\documentstyle[11pt,newpasp,twoside,psfig]{article}
\index{pulsar wind nebulae!Crab Nebula}

\def\cxo{{\em Chandra}}

\def\edcomment#1{\iffalse\marginpar{\raggedright\sl#1\/}\else\relax\fi}
\reversemarginpar

\begin{document}
\title{Year-scale morphological variation of the X-ray Crab Nebula}
 \author{Koji Mori, David N. Burrows, George G. Pavlov}
\affil{Department of Astronomy and Astrophysics, 525 Davey Laboratory,
The Pennsylvania State University, University Park, PA 16802, U.S.A.}
\author{J. Jeff Hester}
\affil{Department of Physics and Astronomy, Arizona State University,
Tempe, AZ 85287-1504, U.S.A.}
\author{Shinpei Shibata}
\affil{Department of Physics, Yamagata University, Yamagata 990-8560, JAPAN}
\author{Hiroshi Tsunemi}
\affil{Department of Earth and Space Science, Graduate School of
Science, Osaka University, 1-1 Machikaneyama, Toyonaka, Osaka 560-0043
JAPAN}

\begin{abstract}

We present year-scale morphological variations of the Crab Nebula
revealed by {\it Chandra} X-ray observatory. Observations have been
performed about every 1.7 years over the 3 years from its launch.  The
variations are clearly recognized at two sites: the torus and the
southern jet. The torus, which had been steadily expanding until 1.7
years ago, now appears to have shrunk in the latest
observation. Additionally, the circular structures seen to the northeast
of the torus have decayed into several arcs. On the other hand, the
southern jet shows the growth of its overall kinked-structure. We
discuss the nature of these variations in terms of the pulsar wind
nebula (PWN) mechanism.

\end{abstract}

\section{Introduction}

The 5-month monitoring observations of {\it Chandra} and HST showed
short-term (days to weeks) morphological variations of the Crab Nebula
(Mori et al. 2002; Hester et al. 2002). A variable inner ring and wisps
emerging from it visually revealed the existence of the termination
shock of the pulsar wind and its downstream flow, respectively. In
addition to such short-term, therefore more noticeable, variations,
expansion of the outer torus on longer time scale (months) was also
discovered. However, this seems to contradict with the fact that the
angular extent of the torus is almost constant over 25 years (Mori et
al. 2002); if the torus kept expanding at the observed rate, it would
have become almost twice larger during this period.  On the other hand,
the southern jet did not show a strong variation of its overall
structure. This differs from the highly variable jet of Vela (Pavlov et
al. 2003), which is an another famous example of a variable PWN.  Here
we present long-term (years) morphological variations of the Crab Nebula
which are newly discovered comparing {\it Chandra} observations over 3
years.

\section{Observations}

The Crab Nebula has been observed by several PIs with different
objectives. The first observation was performed just after launch
(Weisskopf et al. 2000) and the latest one was done 3.3 years after that
to witness the transit of Titan, Saturn's largest moon, across the Crab
Nebula (Mori et al. 2003, in preparation). In this paper, we used those
two observations as well as one of the monitoring observations which
were performed almost at the midpoint of the first and latest
ones. Hereafter, we call them 1st, 2nd, and 3rd epoch observations. They
are roughly spaced at 1.7 years interval. Table~\ref{table:log}
summarizes these observations.

\begin{table}
\caption{Observational logs of {\it Chandra} observations used in this
paper} \label{table:log}
\begin{tabular}{ccccc}
\tableline
\tableline
Notation & Date & Interval (year)\tablenotemark{a} & Exposure (ksec) & Configuration\tablenotemark{b} \\
\tableline
1st & 08/29/99 & -   & 2.7 & ACIS(3.2)+HETG \\
2nd & 04/06/01 & 1.7 & 2.6 & ACIS(0.2)\\
3rd & 01/05/03 & 3.3 & 35 & ACIS(0.3)+HETG\\
\tableline
 \multicolumn{5}{l}{\footnotesize $^{a}$An interval from the 1st observation}\\
 \multicolumn{5}{l}{\footnotesize $^{b}$Number in parenthesis indicates the frame
 time}
\end{tabular}
\end{table}

\section{Result and Discussion}

In contrast with short-term variations are seen at inner regions, the
long-term variations are prominent at outer regions: the torus and the
southern jet. In the following sections, we will discuss these
variations separately.

\subsection{The northeast of the torus}

\begin{figure}[htb]
\centerline{\psfig{file=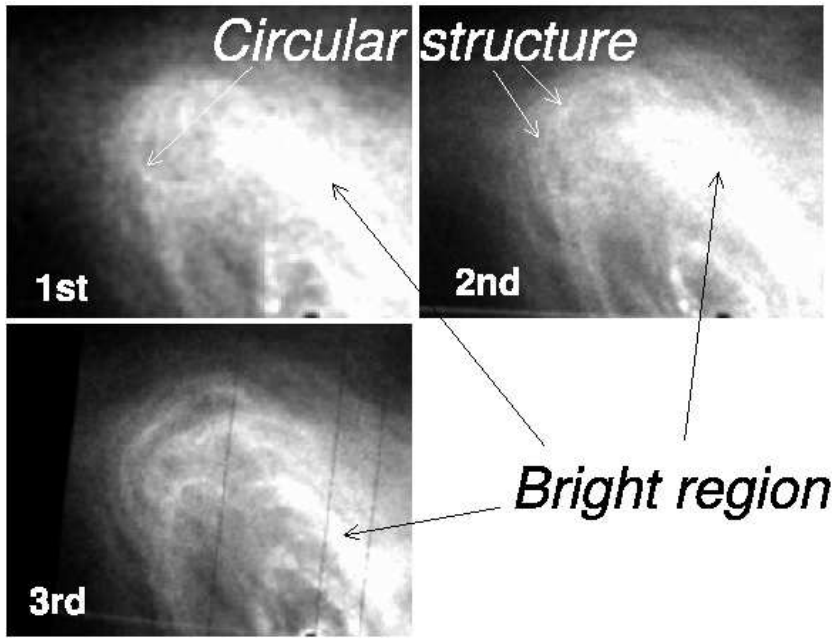,width=11cm}} \caption{Expanded
views around the northeast of the torus of 1st (top left), 2nd (top
right), and 3rd (bottom left) epoch observations. Black lines seen in
the 3rd epoch image are instrumental effects.}  
\label{figure:torus}
\end{figure}

Figure~\ref{figure:torus} shows expanded views of the northeast of the
torus. The brightness scales of these images are normalized to the bright
region of the torus, not by exposure time, because of the differences of
the observational configurations. 

Images of the 1st and 2nd epoch observations are quite similar. The
bright ``thick'' torus is encircled by circular structures at its
northeastern end. As reported previously, these two structures appear to
expand between the 1st and 2nd epoch observations (Mori et al. 2000;
Mori 2002). In contrast with the similarity between the 1st and 2nd
epoch images, the morphological transition from the 2nd to the 3rd epoch
image is remarkable.  The bright region of the torus appears to have
became shrunken and ``thin'', and the circular structures have decayed
into several arcs.

Now it is clear that the torus is not steadily expanding; the overall
extent is almost constant over decades, but the boundary of the torus
varies with an angular scale of a few arcseconds and a time scale of
years. It is as if we were seeing a top of a fountain. Greiveldinger \&
Aschenbach (1999) reported surface brightness variations of the Crab
Nebula using $ROSAT$ HRI observations spanning 6 years. Although $ROSAT$
could not detect a morphological variation, they showed a monotonic
increase of the surface brightness at the northeastern region of the
torus. The brightness variation discovered by $ROSAT$ is most likely
related to the morphological variation discovered by \cxo, suggesting
that the time scale of the variation of the torus might be about a
decade rather than a few years.

\subsection{The southern jet}

\begin{figure}[htb]
\centerline{\psfig{file=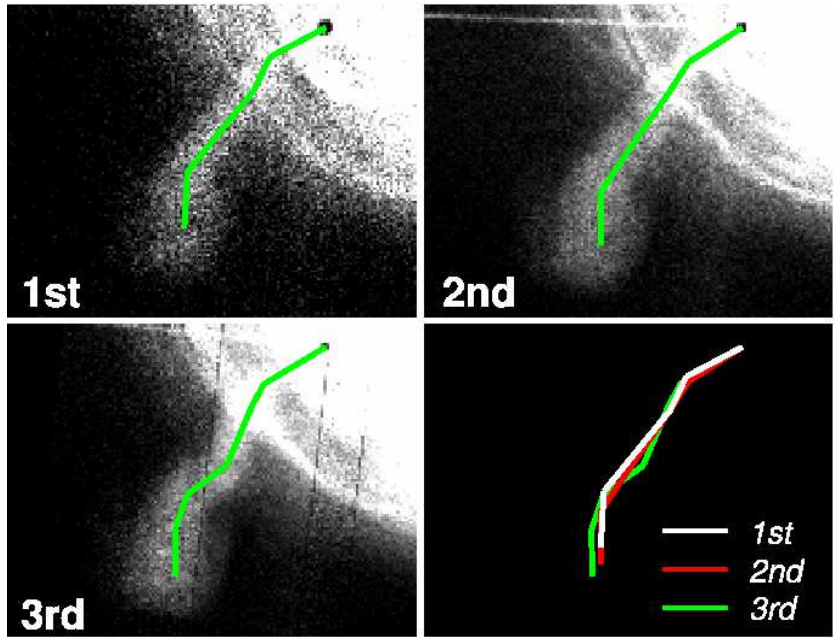,width=11cm}} \caption{Expanded views of
the southern jet of 1st (top left), 2nd (top right), and 3rd (bottom
left) epoch observations. The line in each image traces the axis of the
jet. All three lines are superimposed in the bottom right image.}
\label{figure:jet}
\end{figure}

Figure~\ref{figure:jet} shows expanded views of the southern jet of the
1st, 2nd, and 3rd epoch observations. Although the displacement of the
jet is quite small, the series of images clearly shows the growth of its
overall kinked-structure. The variability of the jet is reminiscent of
northern variable jet of Vela (Pavlov et
al. 2003). Table~\ref{table:jet} compares the time scale of the
variability and the width between the Crab and Vela Jets. The Crab jet
is 10 times larger and varies 10 times slower than the Vela jet. If
these variations are due to MHD instability, the time scale is
proportional to the Alven crossing time, $\tau \sim r/\nu$, where $r$ is
the width of the jet and $\nu$ is the Alven velocity (Begelman
1998). Considering that the Alven velocity in an ultrarelativistic
plasma like the Crab and Vela PWN is more or less the same, above
equation applies to both the Crab and Vela jet. Therefore, it is
suggested that a common mechanism is responsible for the variability of
the Crab and Vela jet.

\vspace{-0.5cm}
\begin{table}[htb]
\begin{center}
\caption{Comparison of variable jets of the Crab Nebula and Vela}
\label{table:jet}
\begin{tabular}{lcc}
\tableline
\tableline
& Time scale (day) & Width (cm)  \\
\tableline
Crab & 150-500 & 2.9 $\times$ 10$^{17}$ \\
Vela & 10-30 & 3 $\times$ 10$^{16}$ \\
\tableline
\end{tabular}
\end{center}
\end{table}

\end{document}